\documentclass[aoas,MSNbibl,nameyear,dvips]{arximspdf}
\usepackage{graphicx}
\doi{10.1214/14-AOAS732} 
\volume{8}
\issue{2}
\pubyear{2014}
\firstpage{1065}
\lastpage{1094}

\makeatletter
\setattribute{abstract}{width}{280pt}
\makeatother

\begin{document}
\begin{frontmatter}

\title{Voxel-level mapping of tracer kinetics in PET studies:
A statistical approach emphasizing tissue life tables\thanksref{T1}}

\thankstext{T1}{Supported in part by the National Institutes of Health
(USA) under
CA-42045 and by Science Foundation Ireland under 11/PI/1027.}

\runtitle{Nonparametric residue analysis}

\begin{aug}
\author[a]{\fnms{Finbarr}~\snm{O'Sullivan}\corref{}\ead[label=e1]{f.osullivan@ucc.ie}\thanksref{m1}},
\author[b]{\fnms{Mark}~\snm{Muzi}\thanksref{m2}\ead[label=e2]{muzi@uw.edu}},
\author[c]{\fnms{David~A.}~\snm{Mankoff}\thanksref{m3}\ead[label=e3]{david.mankoff@uphs.upenn.edu}},
\author[d]{\fnms{Janet~F.}~\snm{Eary}\thanksref{m2}\ead[label=e4]{jeary@uab.edu}},
\author[b]{\fnms{Alexander~M.}~\snm{Spence}\thanksref{m2,T2}}
\and
\author[b]{\fnms{Kenneth~A.}~\snm{Krohn}\thanksref{m2}\ead[label=e6]{kkrohn@u.washington.edu}}

\affiliation{University College Cork\thanksmark{m1}, University of
Pennsylvania\thanksmark{m3}\break and University of Washington\thanksmark{m2}}

\runauthor{O'Sullivan et al.}
\thankstext{T2}{Deceased.}
\address[a]{F. O'Sullivan\\
Department of Statistics\\
University College Cork\\
Cork\\
Ireland\\
\printead{e1}}

\address[b]{K. A. Krohn\\
M. Muzi\\
A. M. Spence\\
Department of Radiology\\
University of Washington\\
Seattle, Washington 98195\\
USA\\
\printead{e6}\\
\phantom{E-mail:\ }\printead*{e2}}

\address[c]{D. A. Mankoff\\
Nuclear Medicine\\
\quad and Clinical Molecular Imaging\\
University of Pennsylvania\\
Philadelphia, Pennsylvania 19104\\
USA\\
\printead{e3}}

\address[d]{J. F. Eary\\
Department of Radiology\hspace*{21.5pt}\\
University of Alabama\\
Birmingham, Alabama\\
USA\\
\printead{e4}}
\end{aug}

\received{\smonth{2} \syear{2010}}
\revised{\smonth{12} \syear{2013}}

%
\begin{abstract}
Most radiotracers used in dynamic positron emission tomography (PET)
scanning act in a linear time-invariant fashion so that the
measured time-course data are a convolution between the
time course of the tracer in the arterial supply and the local tissue
impulse response, known as the tissue residue function. In statistical
terms the residue is a life table for the transit time of injected
radiotracer atoms.
The residue provides a description of the
tracer kinetic information measurable by a dynamic PET scan.
Decomposition of the residue function allows separation of rapid
vascular kinetics from slower
blood-tissue exchanges and tissue retention.
For voxel-level analysis, we propose that residues be modeled
by mixtures of nonparametrically derived basis residues obtained by
segmentation of the full data volume. Spatial and
temporal aspects of diagnostics associated with voxel-level model
fitting are
emphasized.
Illustrative examples, some involving cancer imaging studies, are presented.
Data from cerebral PET scanning with $^{18}$F fluoro-deoxyglucose (FDG)
and $^{15}$O water (H2O) in normal subjects is used to evaluate the approach.
Cross-validation is used to make regional comparisons between residues
estimated using adaptive
mixture models with more conventional compartmental modeling techniques.
Simulations studies are used to theoretically examine mean square error
performance and to explore the benefit of
voxel-level analysis when the primary interest is a statistical summary
of regional kinetics.
The work highlights the contribution that multivariate analysis tools
and life-table concepts can make
in the recovery of local metabolic information from dynamic PET
studies, particularly ones in
which the assumptions of compartmental-like models, with residues that
are sums of exponentials, might not be
certain.
\end{abstract}



\end{frontmatter}

\section{Introduction}\label{sec1}

Dynamic PET studies provide the opportunity to image functional
metabolic parameters of tissue in-vivo [\citet{phelps00}]. Although there
have been many developments in this direction [e.g., \citet{cunn-jones}, \citet{murase}, \citet{osull93},
Muzi et al. (\citeyear{Muzi12}),
Veronese et al. (\citeyear{VRT})], no procedure has yet been\vadjust{\goodbreak}
widely adopted for routine use. Most often quantitation of dynamic PET studies
is based on consideration of a single time point for
a user-defined region of interest (ROI). In view of the complexity of
PET imaging and its expense, this is unsatisfactory.
As most radiotracers used in PET act in a linear and time-invariant fashion,
dynamic PET imaging measures the convolution between the activity of
the tracer in the arterial blood supply and
the tissue impulse response. The impulse response is known as the
tissue residue function.
In statistical terms the residue is the life table associated with
the collection of PET tracer atoms introduced, typically by intravenous
injection, to the circulatory system.
The residue has its roots in the seminal indicator dilution work of
\citet{MZ54}.
Kinetic analysis of PET data is substantially concerned with modeling
and estimation of the residue function.
To this end, there are a suite of commonly used compartmental models
[\citet{huang86}]. However, while compartmental models
adequately represent
the biochemistry of well-mixed homogeneous in-vitro samples,
they are not necessarily well suited to represent micro-vascular flows
and micro-heterogeneity that are part of in-vivo tissue [Bassingthwaighte
(\citeyear{Bass70}), Li, Yipintsoi and Bassingthwaighte (\citeyear{li}) and \citet{ostergaard}].
Consequently, there is interest in more flexible nonparametric
approaches to the estimation of the tissue residues.
Among the most popular approaches is the spectral method introduced by
\citet{cunn-jones}. Here residues are approximated
by nonnegative sums of exponentials, whose amplitudes and rate
constants are adapted to the data;
see Veronese et al. (\citeyear{VRT}) for a recent treatment and review.
Spectral methods have the complexity of requiring estimating of a set
of intrinsically nonlinear exponential rate constants.
This is a significant practical computational challenge; see Zeng et al. (\citeyear{zeng}), for example.
But spectral methods also have a theoretical limitation in that they
force the negative-derivative of residue function, aka
the transit time density of tracer-atoms, to be monotonically
decreasing from a mode at zero.
This assumption is at odds with micro-vascular flow measurements which
support a more log-Normal or Gamma-like form for the transit time density.
If the residue is to be estimated nonparametrically,
it is desirable to have a procedure, like that given in Hawe et al. (\citeyear{HaweOS}) or O'Sullivan et al. (\citeyear{Sull09}), that does not impose an
unrealistic physiologic assumption on the
residue function {ab initio}.

Our focus here is on voxel-level estimation.
The method approximates voxel-level residues by a mixture of \textit{basis} residue functions that have been
optimized by applying a backward elimination technique to a
segmentation of the entire volume of data.
The use of mixtures in this setting is not new [\citet{osull93}],
however, unlike the previous work, which has involved approximation of
mixtures of compartmental models by
a compartmental model form, the current approach does not require this step.
An important aspect of the methodology is decomposition of the tissue
residue to separately focus on
characteristics associated with short transit times of tracer atoms in
the vasculature, distinct from slower transit times associated with
blood-tissue exchange and retention.
This decomposition parallels the often separate consideration given to
early and
late life-time mortality patterns in human life tables.
The methodology leads to a practical quadratic programming-based
algorithm for voxel-level residue
reconstruction and associated generation of functional metabolic images
of parameters of interest.
For a typical dynamic PET study the analysis, including the
segmentation steps, runs on a 3.2~GHz PC in less than
30 minutes.
In the context of PET scanning in cancer applications, that is, about
90\% of all clinical PET imaging studies,
this is completely adequate for routine operational use.

Section~\ref{sec2} presents the basic statistical models underlying the approach.
Inference and model selection methodology are developed in Section~\ref{sec3}.
Section~\ref{sec4} presents illustrations with imaging data from both normal
subjects and cancer patients.
Performance of the methodology for
$^{18}$F fluoro-deoxy\-glucose (FDG) and $^{15}$O water (H2O) imaging
studies is considered in Section~\ref{sec5}. This includes comparisons with
compartmental model
analysis and more
theoretical evaluation via simulations.

%

\section{Theory}\label{sec2}

Let $C_{T}(t,x)$ represent the concentration of tracer atoms at time
$t$ in a tissue
voxel with three-dimensional spatial coordinate $x$. $C_{T}(t,x)$,
measured as activity per unit mass (mg) of tissue, evolves in
response to the localized arterial input function, denoted $C_{P}(t,x)$
and measured as activity per unit volume (ml) of whole blood.
The basic assumption of most PET imaging is that
the interaction of the tracer with the tissue
can be approximated as a linear and time-invariant process.
Thus, the measurable concentration arises as a convolution between
the tissue response and the arterial input function
%
\begin{equation}\label{eq1}
C_{T}(t,x)=\int_{0}^{t}R(t-s,x)C_{P}(s,x)\,ds.
\end{equation}
Here $R (t,x)$ is tissue response and,
borrowing terminology of the work of Meier and Zierler
(\citeyear{MZ54}) on indicator dilutions, $R(\cdot,x)$ is called the residue function.
Formally, $R(\cdot,x)$ is proportional to the impulse response of the
tissue at location $x$ and
has units of flow (ml$/$g$/$min). If all tracer atoms were instantaneously
introduced in a unit volume of blood,
$R(\cdot,x)$ would give the number remaining in the tissue as a
function of time.
If $N$ tracer atoms per ml are introduced in the arterial blood supply
to the tissue, for small time increments $\Delta t$,
then the number of those atoms, per gram of tissue, remaining in the
tissue over the time interval $[t,t+\Delta t]$ is $N R(\cdot,x)\Delta t$.
In statistical terms, $R(\cdot,x)$ defines the life table for the time
variable which measures the duration of stay of the tracer atom in the
tissue---the transit time.
Tracer atom transit times are viewed as random---they arise from a
range of interactions with an array of micro-vascular flow paths,
transporters, enzymes, ligands, etc.
that have the potential to influence the overall length of time the
tracer atom is in the local tissue region.
Measurable transit times are restricted to the observation window $[0,
T_e ]$ of the scanning and, apart from the complexity of indirect measurement
by the convolution equation, there is censoring because tracer atoms
decay over time.
Such restrictions will be familiar, as they arise in traditional
life-table work.

\subsection{Decomposition of residue and key summary parameters}\label{sec2.1}
To better understand the residue, it is helpful to
separate the early (vascular) component from the later components that
are associated with longer term interaction with the tissue and also
retention.\setcounter{footnote}{2}\footnote{This is similar to how a mortality life table might
be dissected to focus on survival patterns at different
stages of life.}
Using $\tau_v \in( 0, T_e)$ as a cutoff for rapid (large-vessel)
transit times, a decomposition of the residue is obtained as
%
\begin{equation}\label{eq2}
R(t,x) =R_B (t,x) + R_D (t,x) + R_X
(t,x), 
\end{equation}
where $R_B (t,x) = R(t\land\tau_v,x )$, $R_D (t,x)=R(t\lor\tau_v,x )
- R(T_e , x)$ and $R_X (t,x)$ is the constant $R(T_e, x)$.
We refer to $R_B$ as the rapid vascular component, $R_D$ as the
exchangeable or in-distribution component and $R_X$ as the (apparent)
extracted component. \textit{Apparent} is used because the ultimate
(asymptotic) extraction is not strictly observable based on the
finite duration of the study, however, as it is common to choose $T_e$
large enough that there would be little further decline in the residue
at times greater than $T_e$, $R(T_e, x)$ should be a good approximation
to the relevant flux of the tracer atoms into tissue.
The decomposition in (\ref{eq2}) is dependent on the value of $\tau_v$ (and
$T_e$). For human imaging, the choice of $\tau_v=1$ minute is
reasonable, as this matches
the early vascular distribution time for intravenously injected
contrast agents, upon which the standard scanning duration
used to assess local blood volume parameters in computerized tomography
(CT) and magnetic resonance (MR) is based [Provenzale et al. (\citeyear{AJNR})].
In the absence of other information, the temporal resolution of a PET
study for the residue can be
no better than the temporal sampling of scanning and the sharpness of
the arterial input resulting from the intravenous injection of the tracer.

Each component of the residue decomposition in (\ref{eq2}) is itself a residue
or life table.
The extracted component is constant but the vascular and distribution
components carry information beyond scale.
Key parameters for a residue function are its maximum and integral
values, which represent the flow and volume occupied by the collection
of tracer atoms defined by the residue [Meier and Zierler
(\citeyear{MZ54}), Hawe et al. (\citeyear{HaweOS})].
So based on the decomposition in (\ref{eq2}), we identify five summary
parameters of particular interest---vascular flow and volume ($K_B (x)
$, $V_B (x)$), distribution
flow and volume ($K_D (x)$, $V_D (x)$) and the apparent flux ($K_i
(x)$) which is seen as the net flux of tracer into tissue up to time $T_e$.
A further parameter of interest is the extraction fraction, defined by
$ \zeta(x) ={K_i (x)}\cdot{(K_B (x) + K_D (x) + K_i (x))^{-1}}$.
In the case where the residue is exponential, for example, a
1-compartment model [e.g., Bassingwaighte (\citeyear{Bass70})] with rate constants
$K_1$ and $k_2$, for $\tau_v=0$, the flow $K_D$ reduces to $K_1$ and as
$T_e \rightarrow\infty$ the exchangeable volume $V_D \rightarrow\frac
{K_1}{k_2}$.
Also,
for a 2-compartment FDG model [Phelps et al. (\citeyear{Phelps79})] with $k_4=0$
and $\tau_v=0$, as $T_e \rightarrow\infty$ $K_D \rightarrow\frac
{K_1k_2}{k_2+k_3}$,
$V_D \rightarrow\frac{K_D}{k_2+k_3}$ and
the flux value $ K_i (x) \rightarrow\frac{K_1 k_3}{k_2 + k_3}$.\footnote{Consideration of asymptotic quantities might be criticized, as it
assumes the underlying physiology is constant---extrapolation beyond
the observation window is always fraught with difficulty.}\vspace*{1pt}

Substitution of the residue decomposition (\ref{eq2}) into (\ref{eq1}) gives a
decomposition of the tracer tissue concentration as a sum of vascular
($C_V$), in
distribution ($C_D$)
and extracted ($C_X$) components
%
\begin{equation}\label{eq3}
C_{T}(t,x)=C_{V}(t,x)+ C_D (t,x) +
C_X (t,x).
\end{equation}
The sum $C_E(t,x)=C_D (t,x) + C_X (t,x)$ is the extravascular component.
Examples of this are shown in Section~\ref{sec4}.
As $R_X$ is constant, $C_X (t,x)$ is the product of flux and the
cumulative arterial activity.
At late time points,
vascular and exchangeable concentration are safely ignored so the late
time concentration is effectively proportional to
the cumulative arterial activity.
This is the basis of a model-free approach to the analysis of flux
[Patlak, Blasberg and Fenstermacher (\citeyear{Patlak})].

\subsection{Additive modeling of the tissue residue}\label{sec2.2}

A variety of blood-tissue exchange models, for example, Bassingwaighte
(\citeyear{Bass00}), Huang and Phelps (\citeyear{huang86}) and Gunn et al.
(\citeyear{Gunn}), as well as
many general life-table methods, for example, Cox and Oakes (\citeyear{Cox01}) and
\citet{lawless}, might be used to approximate tissue residue functions.
We should allow
any approach that does not systematically misrepresent the
physiologic/metabolic processes involved.
Validation of model formulations for PET tracers is difficult. In-vitro studies
clarify important biochemical transformations involved, but satisfactory
in-vivo validation of model assumptions related to the structure of
micro-vasculature flows and heterogeneities is not possible.
The most widely used one- and two-compartment models in PET reduce to
representation of the residue by sums of mono-exponential functions.
While these models may adequately represent the biochemistry involved,
their ability to describe the complexities of vascular transport is
limited. Indeed, in the
standard
compartmental models the tracer atom transit time density is always
monotonically decreasing, so
the modal transit time for the nonextracted tracer is always zero.
Physiologically this is difficult to justify [Bassingwaighte (\citeyear{Bass70})].

We use an additive model that approximates the local tissue residue by
a positive linear sum of
a fixed set of distinct basis residue functions, $\bar R_1 , \bar R_2 ,
\ldots ,\bar R_J$, that have themselves been derived from a nonparametric
analysis of time courses
arising from a full
segmentation of the data volume.
The model is
%
\begin{equation}\label{eq4}
R(t,x)\approx\alpha_{1}(x) \bar R_{1}(t)+
\alpha_{2}(x) \bar R_{2}(t)+\cdots+\alpha\label{eq:rr}
_{J}(x) \bar R_{J}(t),
\end{equation}
where the $\alpha_{j}$'s are nonnegative constants. For simplicity, the
basis residues are normalized to have maximum of unity, that is, $\bar
R_j (0)=1$ for $j=1,2,\ldots,J$.
Assuming $C_P (t,x)$ can be described
by a delayed version of a sampled
arterial time-course $C_P (t)$, which, in view of the temporal
resolution of PET, is reasonable, equation (\ref{eq4}) implies
%
\begin{eqnarray}\label{eq5}
C_{T}(t,x)&=& \alpha_{1}(x) \bar C_{1} \bigl(t-
\Delta(x) \bigr)+\alpha_{2}(x) \bar C_{2} \bigl(t-\Delta(x)
\bigr)+\cdots
\nonumber
\\[-8pt]
\\[-8pt]
\nonumber
&&{}+ \alpha_{J}(x) \bar C_{J} \bigl(t-\Delta(x)
\bigr),
\end{eqnarray}
where $\bar C_j (t) = \int_0^t \bar R_j (t-s) C_P (s) \,ds$ for $j=1,2,
\ldots, J$.
For known delay, $\Delta(x)$, the model is linear in the $\alpha$-coefficients.
Note estimation of the $\alpha$'s in (\ref{eq5})
allows the local residues to be determined by equation (\ref{eq:rr}); from
them associated flow and volume parameters of Section~\ref{sec2.1} can be recovered.

\subsubsection*{Remarks}
1.
If the $\bar C_j$'s in (\ref{eq5}) correspond to specific regional
time courses,
a~mixture model interpretation for the model can be developed.
This is reasonable, as the population of available metabolic pathways
for a tracer atom is determined by the
profile of enzymes, receptor ligands or transporters that are represented.
Across a collection of voxels these profiles vary with a greater
representation of certain characteristics in some voxels than in
others. Thus,
the transit time for a randomly chosen tracer atom in voxel $x$ can be
expected to select a metabolic pathway in accordance with
the distribution of pathways available within the voxel, and the
$\alpha$-coefficients (scaled to sum to unity) could be
viewed as a set of mixing proportions; see O'Sullivan (\citeyear{osull93}, \citeyear{osull06}).
The form in equation (\ref{eq:rr}) can also be viewed as an example
of a general multivariate factor analysis (without reference to
residues). Such models have been used to describe PET time-course data;
see, for example,
Kassinen et al. (\citeyear{kaasinen}), \citet{layfield}, Lee
et al. (\citeyear{lee}) and Zhou
et al. (\citeyear{zhou}).

2. A tissue region can contain significant nonarterial blood vessels.
Depending on tissue location,
separate signals associated with
major blood pools in the circulatory system, such as the right
ventricle of the heart, the lungs, venous blood and the venous supply path
from the injection site to the heart, might need to be considered.
This can be accomplished by augmenting equation (\ref{eq5}) to
include terms representing
nonarterial blood signals. Obviously this is particularly relevant in
the thoracic imaging where the direct or indirect (via a spillover artifact)
impact of nonarterial cardiac and pulmonary
blood signals can be significant. Venous blood vessels arise
throughout the body, so there is
a case for always including
a venous signal term. But rarely does the simultaneous measurement of
arterial and venous blood activity arise in a PET study.
Venous blood can be viewed as a response to the arterial supply, so
the venous
signal is sensibly represented as a whole-body response to the
arterial supply---$C_V (t) = \int_0^t R_V (t-s) C_P (s) \,ds$.
Thus, if an explicit venous signal is not available, the structure of
our modeling approach allows for the component $R_j$-residues to adapt to
$R_V$ so that the overall tissue residue will have the venous
component included.
As mean transit times from arterial to venous blood are short ($<$1
minute), our proposed decomposition of the residue with $\tau_b <1$
will be a combination of pure arterial, venous contributions. Hence,
$V_B (x)$ should be viewed as an estimate of the volume (per mg) of
large arterial and large
venous vessels in the tissue. Thus, if an explicit venous blood signal
is included ($\alpha_{0}(x) C_{V}(t-\Delta(x))$),
the local estimate of blood volume should be the sum of the venous
volume [i.e., $\alpha_0 (x)$]
and $V_B(x)$ from the estimated residue in~(\ref{eq2}).

3.
Due to limited resolution, voxel-level data are subject to mixing and
partial volume effects, which are reasonably modeled by mixtures; see
also Section~\ref{sec3.3}.

\section{Inference techniques}\label{sec3}

The estimation of voxel-level residues involves three steps.
First, a segmentation procedure is applied to extract scaled time-course
patterns from the measured set of voxel-level time courses in the data.
Next, the time courses are analyzed to recover a nonparametric residue
function for each and a backward elimination procedure is used to
obtain a reduced set of
basis residue functions. The final step
does voxel-level optimization of $\alpha$-coefficients and delay
in equation (\ref{eq5}) with subsequent evaluation of the voxel
residue in equation (\ref{eq:rr}) and the key parameters identified
in Section~\ref{sec2.1}. The details involved in each of these
steps are presented below. As the analysis is based on a voxel-level fitting
process, the residuals associated with the fitting process provide
useful diagnostic information.
Some proposals for examining the temporal and spatial patterns in those
residuals are
indicated.

\subsection{Segmentation}\label{sec3.1}

A split-and-merge segmentation procedure from\break \citet{osull06} is
used. The procedure groups voxels on the basis of
the shape of the measured time course, the \textit{scaled} time course.
The splitting employs a principal component analysis to recursively
divide the
tissue volume into a large collection (typically 10,000) of
hyper-rectangular regions
whose {scaled} time-course patterns show maximal homogeneity. The merging
process then recursively combines (initially with a constraint to
ensure that segments consist of
contiguous collections of voxels) these regions to create a collection of
regions with high average within-region homogeneity. For the analyses
reported here, the number of segments used is taken to be large enough
to explain 95\% of the variance in the {scaled} time-course
data, about 7--12 segments for a typical cerebral study and 15--20
segments for a chest or abdominal study.
The choice of the number of segments was examined in \citet{osull06}.
As the focus of the algorithm is on \textit{scaled} time-course
information, the extracted
segments are well suited for use in subsequent mixture modeling. A particular
advantage of the scaled approach is that it results in fewer segments
than might otherwise be required to explain a comparable proportion
of variance in the data.
It is helpful to display segments to connect with anatomy.
If the average scaled time course for a segment is given by a vector
$\mu$,
imaging the $\mu$-weighted average of the voxel-level time-course data
in the
segment is effective. Some examples are shown in Section~\ref{sec4}.
%

\subsection{Construction of basis residues}\label{sec3.2}

The time bins of data acquisition are $[$\b{t}$_{b},\bar{%
t}_{b})$ for $b=1,2,\ldots,B$.
The segmentation algorithm provides a mean time course and associated
sample variance vector, $\{(y _{kb},v_{kb})$ for $b=1,2,\ldots,B\}$, for
each of $k=1,2,\ldots,K$ segments. Arising from the underlying Poisson
emissions that are the basis of the imaging process, regional
time-course data have an approximate quasi-Poisson structure [see
\citet{Hues84}; Carson et al. (\citeyear{carson}) and \citet{HaweOS}], that is,
%
\begin{equation}\label{eq6}
\mathrm{E} ( y_{kb} )= \mu_{kb}     \quad\mbox{and}\quad
\mathrm{V} (y_{kb} )\approx\phi_k \mu_{kb}
\end{equation}
for $\phi_k>1$.
Thus, up to known calibration factors (incorporating time-bin duration
and voxel dimension), the mean values are proportional to the integrated
concentration per milligram of radioactive tracer atoms which in turn
is a function of the regional residue $R_k$ and input arterial supply:
%
\begin{equation}\label{eq7}
\mu_{kb} = \int_{\underline{t}_b}^{{\bar t}_b}
C_{k}(t) e^{-\lambda t} \,dt     \quad\mbox{and} \quad    C_{k}(t) =
\int_{0}^{t}R_{k}(t-s)C_{P}(s-
\Delta_{k})\,ds, \label{eq:mu}
\end{equation}
where $\lambda$ represents isotope decay and $\Delta_k$ is an
appropriate delay. Note $C_{k} (t)$ here represents
the total (radioactive and nonradioactive) tracer atom concentration at
time $t$.
An initial set of possible residue basis elements are obtained by
representation of the $R_k$'s
in terms of B-splines [O'Sullivan et al. (\citeyear{Sull09})].
Here the B-spline coefficients and the delay $\Delta_k$ are
optimized by weighted least squares with weights $\varpi_{kb}$ given by
$v_{kb}^{-1}$.

We seek to express the segment residues in terms of a reduced set of basis
residue forms. Since retention is apparent in nearly all residues, it
makes sense to ensure that the
constant residue function (a \textit{Patlak} term, cf. Section~\ref{sec2.1}) is a
fixed member of any basis. Thus, the focus of the basis search is on
representation of the nonretained residue components.
Suppose we have a set of $J-1$ such normalized basis elements denoted
$\bar R_j$ for $j=1,2 \ldots, J-1$ as well as the constant unit value
Patlak residue, $\bar R_J$. With these, the residue for the $k$th
segment ($R_k$) is approximated by the (nonnegative) linear combination
%
\begin{equation}\label{eq8}
R_k \approx\beta_1 \bar R_1+
\beta_2 \bar R_2+ \cdots+\beta_J \bar
R_J \label{eq:rb}.
\end{equation}
By substitution into (\ref{eq:rb}), the $\beta$-coefficients and delay
can be optimized to the observed segment data.
A weighted least squares fitting is used with weights given by $\varpi
_{kb}$. An approximate
unbiased risk assessment criterion is used to obtain an overall
assessment of the $J$-component basis.
The target loss is the weighted square error difference between the
true segment mean vectors and their estimated values based on
the $J$-component approximation
%
\begin{equation}\label{eq9}
L(J) = \sum_{kb} \varpi_{kb} \bigl[
\mu_{kb}-\hat\mu^{(J)}_{kb} \bigr]^2.
\end{equation}
With $z_{kb}=\sqrt{\varpi_{kb}} y_{kb}$ and $\hat z_{kb}=\sqrt{\varpi
_{kb}} \hat\mu^{(J)}_{kb}$,
\[
z_k' \hat z^{(J)}_{k} = \sum
_b \varpi_{kb} y_{kb} \hat
\mu^{(J)}_{kb} = \sum_b
\varpi_{kb} \mu_{kb}\hat\mu^{(J)}_{kb} +
\sum_b \varpi_{kb}( y_{kb}-
\mu_{kb} ) \hat\mu^{(J)}_{kb} .
\]
So in the case that the vector $\hat\mu^{(J)}_{k} $ is a linear
function of the weighted data, that is, $\hat z^{(J)}_{k}=H_{J} z_k$,
$\mathrm{E} (z_k' \hat z^{(J)}_{k}) = \mu_k' \mathrm{E} (\hat
z^{(J)}_{k}) + \operatorname{trace}  [ H_J \Sigma_k  ]$,
where $\Sigma_k$ is the covariance of $z_k$. But separate time frames,
which involve distinct emission events, are statistically
independent, so $\Sigma_k$ is diagonal. From equation (\ref{eq:mu}),
the diagonal entries of $\Sigma_k$ are approximately $ \phi\varpi_{kb}
\mu_{kb}$ for $b=1,2, \ldots, B$.
Letting $\hat D_k$ be the diagonal matrix with entries $\varpi_{kb}
y_{kb}$, we are led to
the criterion
\[
C(J) = \sum_k \biggl\{\sum
_b \varpi_{kb} \bigl[y_{kb}-\hat
\mu^{(J)}_{kb} \bigr]^2 + 2 \phi_k
\operatorname{trace} ( H_J \hat D_k ) \biggr\}
\]
for evaluation of the loss $L(J)$ in equation (\ref{eq9}). Since $E C(J) =
E(L(J)) + \phi\sum_{kb} \varpi_{kb} \mu_{kb}$,
up to a constant that is independent of $J$, $C(J)$ is an unbiased
estimator for $L(J)$.
With $\phi_k=1$, this statistic has a close connection to the Rudemo
and Bowman
cross-validation statistic for density estimation; see, for example,
O'Sullivan and Pawitan (\citeyear{OSPA96})
who gave a corresponding risk estimator for bandwidth selection in
Poisson deconvolution problems.
For practical use a choice of $\phi_k$ must be made---the ratio $\hat
\phi_k = \sum_b v_{kb}/\sum_b y_{kb}$ seems reasonable for this.
If $H_J$ is a projection, that is, $\hat\mu^{(J)}_{k} $ is
approximately obtained by a weighted least squares
regression, $\operatorname{trace} (H_J) = J$ and $\operatorname{trace}  ( H_J \hat D_k  )$
might reasonably be approximated as $J \bar d_k$,
where $\bar d_k = \frac{1}{B}\sum_b\varpi_{kb} y_{kb}$. Thus, we obtain
a criterion with a familiar AIC-like [\citet{Akaike}] form\footnote{It should be clear
that this is not in fact an information criterion in the formal sense---only
first and second order moment
assumptions are involved.}
%
\begin{equation}\label{eq10}
\hat C(J) = \sum_k \biggl\{\sum
_b \varpi_{kb} \bigl[y_{kb}-\hat\mu
^{(J)}_{kb} \bigr]^2 + 2 \hat\phi_k
\bar d_k J \biggr\}.
\end{equation}
Using this criterion, a backward elimination procedure is used to
select a subset of the starting $K+1$ basis residues (corresponding to
the segment data residue fits plus the Patlak residue).
The resulting basis set is applied in the analysis of voxel-level data.

\subsection{Voxel-level optimization and local averaging}\label{sec3.3}

At the voxel-level we have data with substantially the same structure
as the segment-level data, $y_{bk}$ above.
If $ y_{bi}$ for $b=1,2, \ldots,B$ is the PET measured activity at the
$i$th voxel, from equations (\ref{eq6}) and (\ref{eq7}) we have
$\mathrm{E} ( y_{ib} )=\mu_{ib}$ and $\mathrm{V} ( y_{ib} )= \phi_i \mu_{ib}$
with
%
\begin{equation}\label{eq11}
\mu_{ib} = \alpha_{i1} \bar\mu_{1b} (
\Delta_i) + \alpha_{i2} \bar\mu _{2b}(
\Delta_i) +\cdots+ \alpha_{iJ} \bar\mu_{Jb}(
\Delta_i) \equiv\bar\mu_{b} (\Delta_i)'
\alpha_i,
\end{equation}
where $\alpha_{ij}= \alpha_j (x_i)$, and with $\bar C_{j}(t) =\int_{0}^{t} \bar R_{j}(t-s)C_{P}(s)\,ds$,
$
\bar\mu_{jb} (\Delta_i) =
\int_{\underline{t}_b}^{{\bar t}_b}
\bar C_{j}(t-\Delta_i) e^{-\lambda t} \,dt$.
This is in the form of a quasi-Poisson regression problem [McCullagh
and Nelder (\citeyear{McCull})]. A standard iteratively reweighted least squares
method is
used for optimization of unknowns. With weights defined by the current
guess, $\omega_{ib} = 1/{\mu_{ib}^{(0)}}$, updated
values for the unknowns are obtained by minimization of
%
\begin{equation}\label{eq12}
\operatorname{WRSS}({\alpha}_i, \Delta_i ) = \sum
_{b=1}^{B}\omega_{ib} \bigl[y_{ib}-
\bar\mu_{b} (\Delta_i)'
\alpha_i \bigr]^{2}
\end{equation}
with positivity constraints on the $\alpha$-coefficients. Thus, for
fixed $\Delta_i$,
fast (exact) quadratic programming codes are used for evaluation of
solutions, making the computation efficient.
Grid search is used for optimization of the delay.
The impact of iterative reweighting on computing time
here is quite minimal: Typically the solutions obtained with a fixed
set of weights, such as
weights which are inversely proportional to the segment variance
($v_{kb}$, if the voxel belongs to segment~$k$),
produce results that are largely
unaffected by subsequent iteration.

\subsubsection*{Local averaging and smoothing}
The residue formulation allows residues for locally averaged (smoothed)
data to be evaluated as
local averages of residue coefficients---if local variation in delay is
negligible.
To see this, suppose $w_{xi}$ for $i=1,2,\ldots,N$ represents a set of
nonnegative weighting coefficients used to
produce the averaged data---$\tilde y_{xb} = \sum_i w_{xi}y_{ib} $ is
the averaged data at voxel $x$.
Assume that the weights are focused, in the sense that $w_{xi} > 0$ for
$i$ in a neighborhood $N_x$ and
the delay variation
is negligible here, that is, $\Delta_i \approx\Delta$ for $i \in N_x$.
Let $\mu_{ib}= \bar\mu_{b} (\Delta)' \alpha_i $ for
$i \in N_x$. From (\ref{eq11}) we have
%
\begin{eqnarray}
\tilde\mu_{xb}&=&\sum_i
w_{xi} \mu_{ib}
\nonumber
\\[-8pt]
\\[-8pt]
\nonumber
&=&\tilde\alpha_{x1} \bar\mu_{1b} (\Delta) + \tilde
\alpha_{x2} \bar\mu_{2b} (\Delta) +\cdots+\tilde
\alpha_{xJ} \bar\mu_{Jb} (\Delta) = \bar\mu_{b} (
\Delta)' \tilde\alpha_x
,
\end{eqnarray}
where $\tilde\alpha_{xj} = \sum_i w_{xi} \alpha_{ij} $.
Thus, the mean of the smoothed data $\tilde y_{xb}$ is given by a
positive linear combination of the basis elements $\bar\mu_{jb} (\Delta)$
in (\ref{eq11}). Using (\ref{eq11}) again, we can also write $\tilde\mu_{xb} =
\int_{\underline{t}_b}^{{\bar t}_b} \tilde C_{x}(t-\Delta_i) e^{-\lambda t}
\,dt$, where
$
\tilde C_x (t) = \int_0^t \tilde R_x (t-s) C_P(s- \Delta) \,ds$ and
%
\begin{equation}\label{eq14}
\tilde R_x (t) = \tilde\alpha_{x1} \bar R_1
(t) + \tilde \alpha _{x2} \bar R_2 (t) +\cdots+\tilde
\alpha_{xJ} \bar R_J (t).
\end{equation}
Hence, the averaged $\alpha$-coefficients define the appropriate local
residues for the averaged image data.
This result has implication for smoothing by local averaging methods.
While data smoothing/filtering
is not the only mechanism used to achieve regularity in statistical
function estimation, it is widely used and very reasonable
in many image processing contexts, including tomography
[\citet{NaWu01}].
The parametric images produced in the examples in Section~\ref{sec4} are
subjected smoothed---the computed $\alpha$-coefficients are convolved
with a 3-D Gaussian kernel whose
coordinate standard deviation is between 1 and 2 times the voxel dimension.

\subsection{Diagnostic evaluation of model adequacy}\label{sec3.4}

To ensure the validity of the
parametric imaging results, we assess the assumptions underlying
the fitted parametric model by examining the temporal and spatial
patterns in the voxel-level standardized residuals:
%
\begin{equation}\label{eq15}
r_{ib}=\sqrt{w_{ib}}\cdot{}[ y_{ib}-
\hat{y}_{ib} ],
\end{equation}
where $w_{ib}=\hat{y}_{ib}^{-1}$.
Both temporal and spatial patterns in these residuals should be examined.
Temporal aspects are best appreciated by plotting uptake data, for
example, time courses
for segments together with fitted modeling results.
The spatial distribution of the residuals can be examined but the
standardization by activity is unstable for low activities (recall that
the raw filtered back-projection reconstructed data can be negative for
some voxels).
Thus, we find it best to use a common set of weights $w_{ib}=\bar w_b$
and compare the spatial pattern of the residual and data
root mean square (RMS):
\begin{eqnarray*}
\mbox{RMS(residual)}&=& \sqrt{\sum_b \bar
w_{b}[ y_{ib}-\hat {y}_{ib} ]^{2}/B}
; \\
\mbox{RMS(data)}&= &\sqrt{\sum_b \bar
w_{b}[ y_{ib}-\bar{y}_{i} ]^{2}/B},
\end{eqnarray*}
where $\bar{y}_{i}$ is the weighted mean of the time course.
Techniques for analysis of temporal and spatial patterns in residuals
are needed, such as estimation of their
spatial covariance structure, which might be related to the PET image
reconstruction
process [Carson et al. (\citeyear{carson}), \citet{maitra}].

\section{Application to real data}\label{sec4}

We now present some case studies to illustrate the parametric imaging
technique.
The first examples come from cerebral imaging in normal subjects using
the two most well-established PET tracers, FDG and H2O.
The second set of examples are from imaging cancer patients using FDG
and H2O as well more experimental tracers.
All studies were carried out on a GE Advance PET scanner at the
University of Washington.
This scanner produces reconstructions of tracer uptake (per time bin of
acquisition)
as an array consisting of $35$ transverse planes/slices and a within-slice
discretization of $128\times128$ voxels.
Slices have a fixed thickness of $4.25$~mm, but the
within-slice voxel resolution can be varied depending on the tissue
volume in the field of view,
typically $2.23$~mm for brain studies to $4.29$~mm for
abdominal and thoracic studies.
The scanner does not use X-ray CT acquisition for attenuation
measurement. Instead
attenuation is derived from PET
transmission scans with a 511 keV rod source. All data were
reconstructed using standard
filtered back-projection methods. Below we present uptake and
parametric images superimposed on images of PET-measured attenuation.
Tracers were injected intravenously and catheterized sampling used to
measure the tracer time course in the arterial blood ($C_P$).

%
%
%
%
%
%
%
%
%

%
%
%
%
%
%
%
%
%

\subsection{Brain studies in normal subjects with FDG and H2O}\label{sec4.1}

\begin{figure}
\centering
\begin{tabular}{@{}c@{}}
\footnotesize{(a) Dynamic 90-minute PET-FDG study in a normal subject}\\

\includegraphics{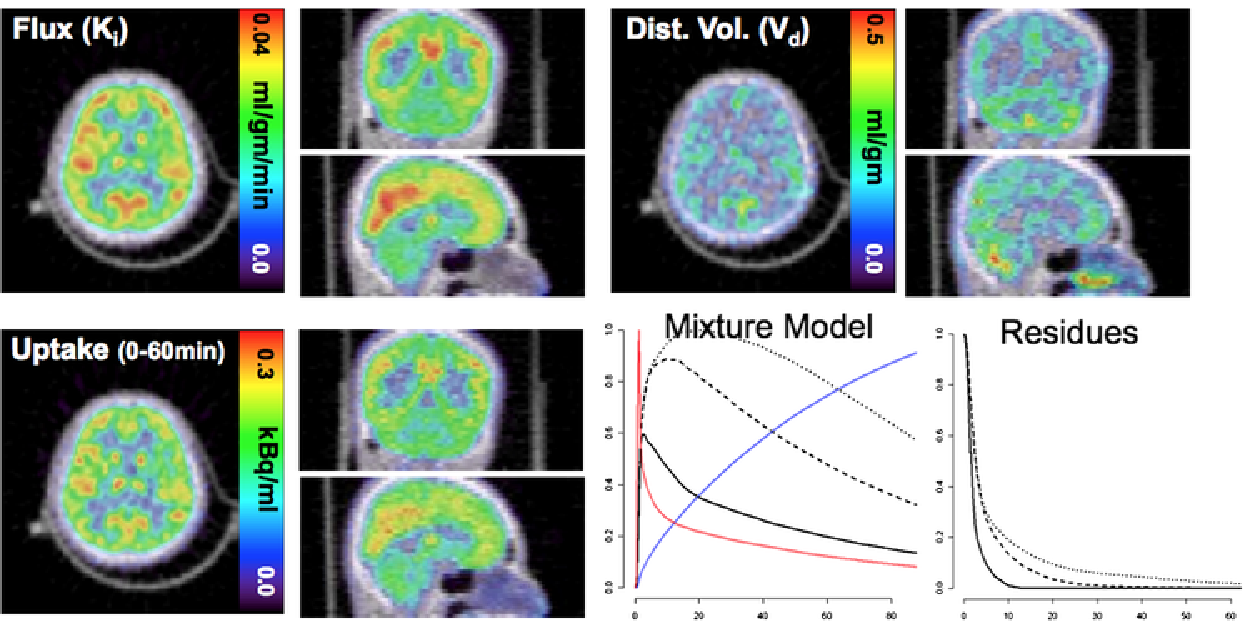}
\\[3pt]
\footnotesize{(b) Dynamic 7.5-minute PET-$H2O$ study in a normal
subject}\\

\includegraphics{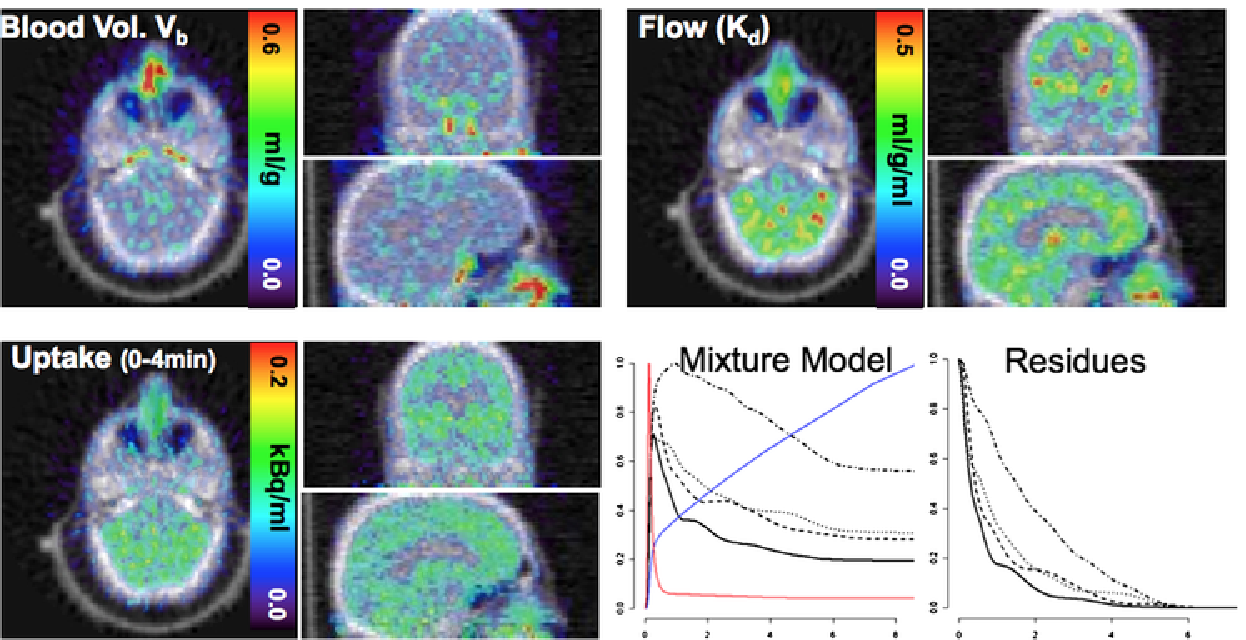}

\end{tabular}
\caption{Color maps of selected kinetic parameters
(cf. Section~\protect\ref{sec2.1})
recovered by residue analysis and the integrated uptake divided by
study duration.
Transverse, sagittal and
coronal slices shown. The gray-scale images give tissue attenuation
measured by the PET transmission scan.
Line plots give
the fitted mixture model [cf. $\bar C_j$ in equation (\protect\ref{eq6})],
including the
arterial input ($C_P$ in red)
and the Patlak component (blue). The normalized, non-Patlak, basis
residues [$\bar R_j$ in equation (\protect\ref{eq5})] are also
shown on the right.}\label{fig:fig1}
\end{figure}

The FDG data set is a series reported in Graham et al. (\citeyear{graham}). For
FDG, dynamic PET imaging was
carried out over a 90-minute time
period ($T_{e}=90$).
The tracer dose (370 MBq)
was injected over a 1-minute period, and
$B=31$ time frames of data were acquired according to the following sampling:
1 minute pre-injection frame followed by 4
15-second frames, 4 30-second frames, 4 1-minute frames, 4 3-minute
frames and ending with
14 5-minute frames.
A 10-region segmentation of the data was used to initialize the mixture
model analysis, that is, $K=10$ in Section~\ref{sec3.2}.
This segmentation explained over 95\% of the variance in the data.
Initialization with $20$ or $40$ segments
produced little or no
change in results.
The final mixture model and associated normalized basis
residues (apart from the constant residue component) had $3$
components; see Figure~\ref{fig:fig1}(a).
Figure~\ref{fig:fig1}(a) shows total tracer uptake as well as computed values of the
flux ($K_i$)
and distribution volume ($V_D$); cf. Section~\ref{sec2.1}, superimposed on the
PET-measured tissue attenuation.
The flux is the most well-resolved parameter here, showing high values
in white
and gray matter structures of the brain. Distribution volume shows a
more diffuse pattern with elevated values
outside the brain (cf. the nasal cavity). In all cases the values of
metabolic parameters are
in agreement with values reported in the literature; cf. Graham et al. (\citeyear{graham}).
Overall, the kinetic analysis of flux and volume of distribution
provides an understanding of
tissue metabolism of FDG that cannot be appreciated from the uptake information.

\begin{figure}[t!]

\includegraphics{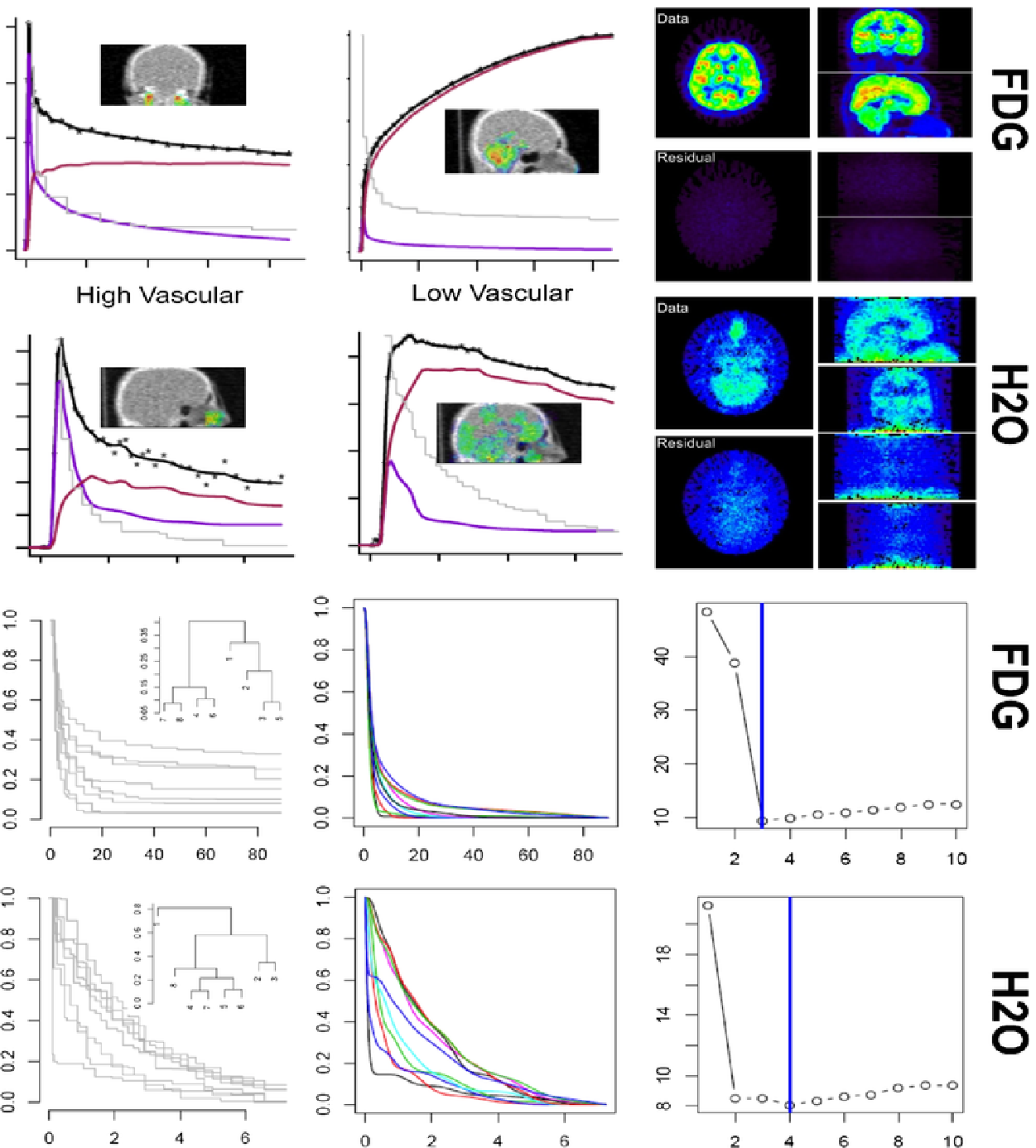}

\caption{{\emph{Top}}: Sample segment time-course data (points) and fitted step-function
nonparametric residues (gray). The fitted residue-model time course
(black line) is shown with vascular (purple) vs. nonvascular (brown)
components: $C_V$ and $C_E$ after (\protect\ref{eq3}). Inserts show the intensity of
the time course within the region of the segment. Right color maps show
the distribution of the RMS for data and residuals from analysis (cf.
Section~\protect\ref{sec3.4}).
{\emph{Bottom}}: (Left) Raw piecewise constant residues for
segments (scaled to unity). A dendrogram obtained by clustering residues
is also provided. (Center) Line plots are scaled B-spline-computed
residues without retention [$\bar R_j$'s in (\protect\ref{eq9})].
(Right) Risk values [$\hat C(J)$ in (\protect\ref{eq10})] plotted against the number of
components ($J$) in the mixture model---optimal number shown with blue
line.}
\label{fig:fig2}
\end{figure}

The H2O data come from a series reported in Muzi et al. (\citeyear{Muzi09}). The
tracer dose (776~MBq)
was injected over a 5-second time period.
Dynamic PET imaging was
carried out over a 7.9-minute time
period ($T_{e}=7.9$).
A~set of $B=42$ time frames of data was acquired according to the
following sampling:
1-minute pre-injection frame followed by 5
3-second frames, 10 6-second frames, 12 10-second frames, 8 15-second
frames and ending with
6 20-second frames.
Similar to FDG, a 10-region segmentation of the data was used to
initialize the mixture model analysis, that is, $K=10$ in Section~\ref{sec3.2}
(values of $K=20$ and $K=40$ were also considered but produced little
or no change in results).
The mixture model and associated normalized basis
residues, apart from the constant residue component, had $4$
components; see Figure~\ref{fig:fig1}(b).
Figure~\ref{fig:fig1}(b) shows total tracer uptake as well as computed values for
vascular blood volume ($V_B$) and the distributional flow component ($K_D$).
Note the vascular blood volume is elevated in the neighborhood of the
internal carodits and the nasal cavity.
The (distributional) flow pattern is elevated in the white
and gray matter structures of the brain. In general, values of
metabolic parameters are
in agreement with those reported in the literature; cf. Muzi et al. (\citeyear{Muzi09}).
Similar to FDG, the kinetic analysis here is seen to add to the
information provided by uptake alone.

Figure~\ref{fig:fig2} presents some diagnostic information for the analyses presented.
Sample segments and the results of the associated nonparametric residue
analysis are shown. The increased variability of H2O is evident. Even
though the injected dose is larger than that of FDG, the shorter
time frames and much shorter half-life of $^{15}$O results in lower
radioactivity per time frame in the H2O study.
Consequently, the data have more noise, as reflected in the displayed
parameters.
The voxel-level residual RMS characteristics of the fit are shown in
Figure~\ref{fig:fig2}.
For FDG, the RMS error is very small relative to the tracer uptake; in
the case of H2O, the RMS error is generally higher
relative to uptake, but
there still does not appear to be a spatially coherent structure to the
lack of fit.
The fitted residue models for the selected segments are in good
agreement with the data; see Figure~\ref{fig:fig2}.
The decomposition of the time course of the fitted models show the
vascular and nonvascular components of the fit [cf. equation (\ref{eq3})].
Note the time courses with high vascular contributions arise from
well-known blood pools in the neck and nasal cavity;
less vascular signals arise from white and grey matter tissue.
The full set of fitted piecewise constant nonparametric residues [Hawe
et al. (\citeyear{HaweOS})] as well as the smoother residues produced by B-spline
fitting [O'Sullivan et al. (\citeyear{Sull09})] are also shown in Figure~\ref{fig:fig2}.
Note these latter residue curves only present the nonretained parts of
the residues, that is,
extraction is subtracted.
The model selection criterion is plotted against the number of
retained components. Three components are
indicated for the FDG and four for the H2O data.
A dendrogram for a hierarchical cluster analysis of the raw residues
(normalized and adjusted for extraction) using a Euclidean distance matrix
shows some
exploratory support for the number of
components selected.
Note that, unlike the model selection statistic, the cluster analysis
does not involve any recursive refitting of the time-course data.
Overall, the analysis and diagnostics are very reasonable.
A more detailed analysis over a series of similar cerebral studies is
provided in the next section.

\subsection{Cancer imaging examples}\label{sec4.2}

The use of PET for metabolic imaging of cancer and its
response to therapy is of current clinical interest [\citet{Abogye},
Krohn et al. (\citeyear{krohn-o}) and Mankoff et al. (\citeyear{mankoff-o})].
While PET-FDG imaging is an indicator of glucose metabolism,
other aspects of cancer biology may also be useful. Cell proliferation, hypoxia,
vascularity, drug resistance and cell death play important roles and the
study of these processes in cancers with PET radiotracers is
being explored with clinical imaging trails [Krohn et al. (\citeyear{krohn-o})].
We present two examples to illustrate applications of our metabolic
imaging procedure in such contexts.
From a methodological viewpoint, these examples demonstrate the
versatility of the residue imaging
technique across a range of radiotracers and also in different target
tissue volumes.

\subsubsection*{I. Brain tumor study}

These data are from a series in Spence et al. (\citeyear{spence09}). The patient
is a 39-year-old with recurrent right parietal anaplastic
astrocytoma.
The patient had follow-up clinical MR scans at 4 and 4.5 years after
initial treatment with a combination of surgery, radiotherapy and chemotherapy.
At the time of follow-up PET scans with
$^{11}C$-labeled acetate (ACE), $^{11}C$-labeled carbon dioxide (CO2)
and $^{18}$F-labeled fluoro-thymidine (FLT) were carried out.
The ACE and CO2 studies were performed at the 4-year follow-up and
the FLT at the 4.5-year follow-up. The focus of these PET scans was to
study the potential for differentiation between tumor regional
radionecrosis (residual but dead tissue) from tumor recurrence. This
issue is difficult to decide on the basis of
the standard clinical MR scan.
Acetate flux is an indicator of membrane lipid synthesis [Yu et al.
(\citeyear{evanyu})] and so may provide an early indicator of tumor recurrence.
Metabolism of acetate
produces carbon dioxide. As a result, the interpretation of acetate is
somewhat confounded by processes associated with the fate of $^{11}C$ in
carbon dioxide. The CO2 PET study is used to address this issue.
FLT has the potential to provide information on the rate of thymidine
utilization and this in turn is
a potential indicator of cell proliferation.

\begin{figure}
\centering
\begin{tabular}{@{}c@{}}
\footnotesize{(a) Brain tumor}\\

\includegraphics{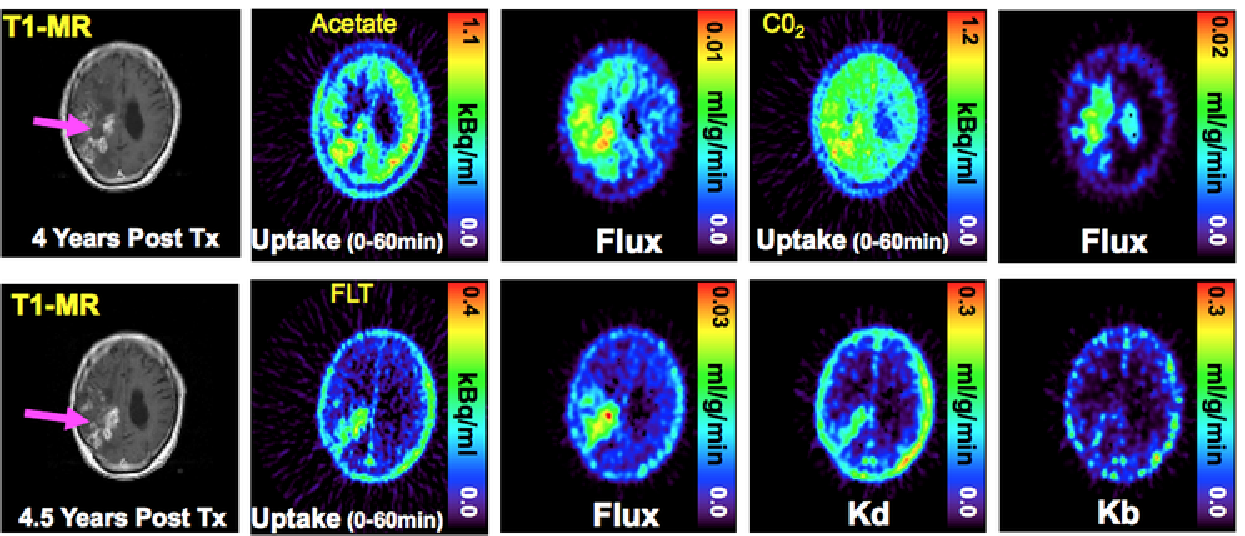}
\\[3pt]
\footnotesize{(b) Breast cancer}\\

\includegraphics{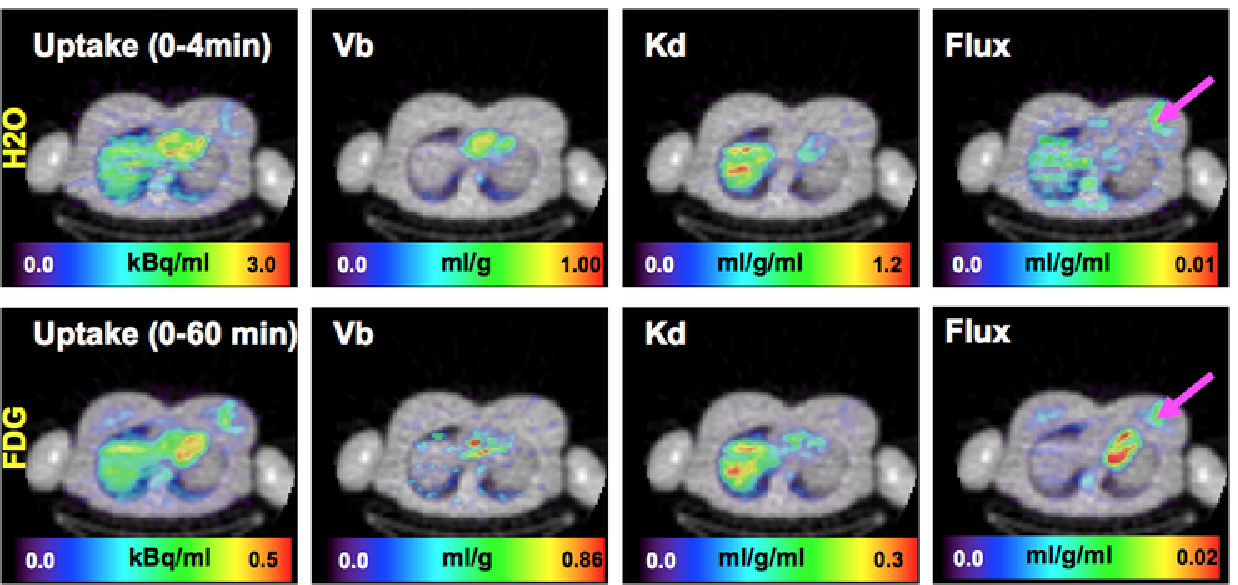}

\end{tabular}
\caption{\textup{(a)} Brain tumor 4 and 4.5 years after treatment. Standard
clinical T1-weighted Magnetic Resonance images---tumor site indicated
with pink arrow.
Dynamic PET imaging with $^{11}C$ Acetate
and $^{11}C$ $CO_2$ at 4-year follow-up were analyzed by residue
analysis to obtain flux estimates.
Dynamic $^{18}F$ fluoro-thymidine PET imaging was analyzed by residue
analysis to obtain flux, $K_d$ and $K_b$ estimates.
Transverse images through a representative plane of the tumor are
shown. \textup{(b)} Patient with locally advanced Breast Cancer imaged with
PET-H2O and PET-FDG.
Transverse slices of tracer uptake and residue calculated metabolic
images of blood and distribution volume and flux.
The grey scale is the measured tissue attenuation. The tumor site
indicated with pink arrows.}\label{fig:fig3}
\end{figure}

Tracers were introduced by 1-minute intravenous infusion.
The sampling protocols for ACE and CO2 were identical: Frame durations
in seconds (some of which were repeated a number of times)
were as follows: 5(18), 10(7), 20(4), 60(4), 180(4), 300(8) for a total
of $B=45$ dynamic scans. With FLT there were $B=47$ frames:
10(10), 20(4), 40(3), 60(3), 120(5), 180(4), 300(18).
An initial $K=10$ region segmentation was used for each data set. The
selected mixture models have 5, 3 and 4 terms for ACE, CO2 and FLT.
In each case, temporal and spatial diagnostics for the mixture model
fit were satisfactory.
Results of metabolic parameters are shown in Figure~\ref{fig:fig3}(a).
The tumor region is apparent on both magnetic resonance (MR) scans. ACE
flux is elevated in parts of this and these areas are also seen as high
FLT flux.
The CO2 flux is elevated in the brain ventricle/choroid plexus. The
choroid plexus makes cerebral spinal fluid (CSF) in part by
carbonic anhydrase to facilitate the exchange between CO2 and water and
bicarbonate, and this starts in the cerebral ventricles.
Thus, it makes physiologic sense that choroid plexus has high CO2 flux.
There is an increased flow of FLT in the skull bone
marrow and that is also elevated (perhaps as a result of
blood-brain-barrier disruption) in the tumor.
Note that the FLT flux is not substantially elevated in the marrow of
the skull here. In normal circumstances this would be unexpected,
however, because
the marrow is lower on the side of the tumor, compared to the other
side, this is likely a radiation effect.
The increased FLT flux in the tumor region abnormality
suggests enhanced cellular proliferation---possible tumor recurrence.
The Spence et al. (\citeyear{spence09}) report found that FLT flux has potential
as an imaging biomarker
for distinguishing proliferation in patients with recurrent gliomas
from radionecrosis, which has a similar MR pattern.
There are a number of National Cancer Institute (NCI) sponsored trials
underway that are investigating the potential of
PET-FLT for guiding the treatment of cancer patients. The quantitation
of flux and flow provided here
enhances the understanding of the information provided
by these studies.

%
%
%
%
%
%
%
%
%
%

\subsubsection*{II. Breast cancer}

This data here is from a series of locally advanced breast cancer
patients studied with PET FDG and H2O
prior to treatment with neoadjuvant chemotherapy.
The study, reported by Dunnwald et al. (\citeyear{dunwald}), showed that the
mismatch between H2O evaluated flow and FDG measured flux, provided
a prognostic indicator of response and disease-free survival.
A bolus intravenous injection of H2O was followed by dynamic PET image
acquisition over a 7.75 minute period. A total of $B=53$ frames were obtained.
A 1-minute pre-injection scan was followed by consecutive frames whose
durations in seconds (with number of repeats)
were as follows: 2(15), 5(14), 10(10), 15(8), 20(5).
A short time later the FDG study was carried out. This entailed
2-minute intravenous infusion of FDG. A 1-minute pre-injection scan was
followed by 24 further frames; these frame durations in seconds
(including repeat number) were as follows: 20(4), 40(4), 60(4), 180(4), 300(4).
There was no blood sampling, but analysis of the right ventricle of the
heart region was used to extract an arterial input function in the analysis.
Additional time courses for venous blood (vena-cava), the pulmonary
veins and the right ventricle of the heart were also recovered from the
PET image data; see \citet{OSull11} for details of the
technique used.
These signals were included
as additional terms in the segmentation and mixture model analysis.
In view of the increased range of tissue structures in the field of
view, a greater number of segments ($K=15$) were used. Selected mixture models
had 5 residue components for FDG and 4 for H2O. In each case temporal
and spatial diagnostics for the mixture model fit were satisfactory.
Parametric images are shown in Figure~\ref{fig:fig3}(b). The raw uptake patterns for
the tracers do not differentiate
between areas with high vascular uptake (heart cavity, liver) and areas
with increased longer term retention (tumor and heart wall).
The vascular and retention patterns are well described by images of
blood volume ($V_B$), distributional flow ($K_D$) and flux ($K_i$).
The increased
apparent retention time of water in the tumor region (seen as a ``flux''
in the water residue analysis) may reflect that less effective
vasculature often is associated with cancers, particular larger
cancers; see Dunnwald et al. (\citeyear{dunwald})
and Jain (\citeyear{Jain05}).
This is an example of a patient whose tumor has low blood flow and high
FDG uptake, which is nicely captured by the analysis.
The greatest FDG flux is in the section of the tumor at the edge of an
apparent necrotic section,
which makes sense as a place where blood flow would be low and FDG
uptake might be high.
We expect the flow measures to be similar from water and FDG
[Tseng et al. (\citeyear{Tseng04})], and the images support this.

%

\section{Analysis of performance}\label{sec5}

We consider cerebral PET studies with FDG and H2O and examine aspects
of the proposed methodology
using both real and simulated data.
One issue is to determine how
regional averages of voxel-level residue estimates compare to
regional residue estimates produced by analysis of regionally
averaged time-course data.
The latter issue is of interest because it is the common way that
regional summaries of PET tracer kinetics are obtained.
We used numerical simulations to explore this matter.
A second issue is whether the approximation capabilities of our residue
basis are adequate for real data.
Recall the underlying mixture model is constructed adaptively by
segmentation of the entire image volume. It is not clear if the derived mixture
basis set has adequate local approximation characteristics.
We compare our mixture modeling approach with an analysis of
regionally averaged time-course data using standard compartment
models.

\subsection{Adaptive mixtures versus compartmental model residues}\label{sec5.1}

As previously indicated, the sample data sets used in Section~\ref{sec4.1} are
taken from two series on normal subjects
reported by Muzi et al. (\citeyear{Muzi09}) and Graham et al. (\citeyear{graham}).
In those series PET time-course data for a number of brain structures
were recovered.
With detailed reliance on co-registered high-resolution MR scans for
anatomic reference,
region of interest (ROI) PET time-course data for a number of brain
structures were created.
For FDG there were 10 regions in each of 12 studies to give a total of
120 ROI data sets. The regions were as follows: cerebellum,
temporal, frontal, parietal and occipital cortex, thalamus, putamen, caudate,
whole brain and white matter.
For H2O, a set of 11 subjects was considered with each subject scanned
twice. In each study between 6 and 9 regions
were considered, giving a total of 184 ROI data sets. The regions
included: choroid, pituitary and salivary gland, ventricle, selected
whole brain regions,
white and gray matter.
In Muzi et al. (\citeyear{Muzi09}) and Graham et al. (\citeyear{graham}), these ROI data
were analyzed using the conventional
compartmental models.\footnote{A description of
the relevant 1- and 2-compartmental models used for H2O and FDG in a
normal brain can be found, for example, in Huang and Phelps (\citeyear{huang86}).}
Here we compare compartmental analysis of these H2O and FDG ROI data with
analysis obtained using the adaptive mixture model recovered using the
segmentation and recursive refinement algorithm of Section~\ref{sec3.1} and
Section~\ref{sec3.2}
of this paper.
The mixture model was then applied to analyze the time-course data for
each of
the ROIs that had been recovered in the study, that is, application of
the model in equation (\ref{eq11}) to the ROI data.
At the same time the ROI data were also analyzed using the relevant
compartmental models.
For reference, the ROI data were also analyzed using a fully
nonparametric procedure with a piecewise constant residue [Hawe et
al. (\citeyear{HaweOS})].
In terms of fit, this gives an effective lower bound on what can be
achieved for a given data set.
Leave-out-one cross-validation residuals [\citet{weisberg}] were
constructed for each analysis method, and the weighted sums of squares
of these residuals
were used for comparisons between different methods.
Absolute cross-validated residuals from the mixture and compartmental approaches
were also subjected to a paired Wilcoxon test [\citet{wilcox}].
The $p$-value for this test provides evidence against the hypothesis of
similarity between the average magnitude of cross-validated residuals
from the compartmental and mixture analysis.
An overall comparison between the compartmental and mixture models is
also carried out. A sign test is applied to the
set of differences between the cross-validated residual sums of
squares for the mixture and compartmental
model. The mixture model is
favored if the percentage of time it outperforms the compartmental
model is significantly greater than 50\%.
We also compare compartmental and mixture model-based estimates of key
kinetic parameters:
flux ($K_i$) and distribution volume ($V_D$) for FDG and blood volume
($V_B$) and the distributional flow ($K_D$) for H2O.
Note in the compartmental case this involves evaluating the fitted
compartmental residue and then using the definitions in Section~\ref{sec2.1}
to evaluate the resultant kinetic parameters.

\begin{figure}
\centering
\begin{tabular}{@{}c@{}}
\footnotesize{(a) FDG data}\\

\includegraphics{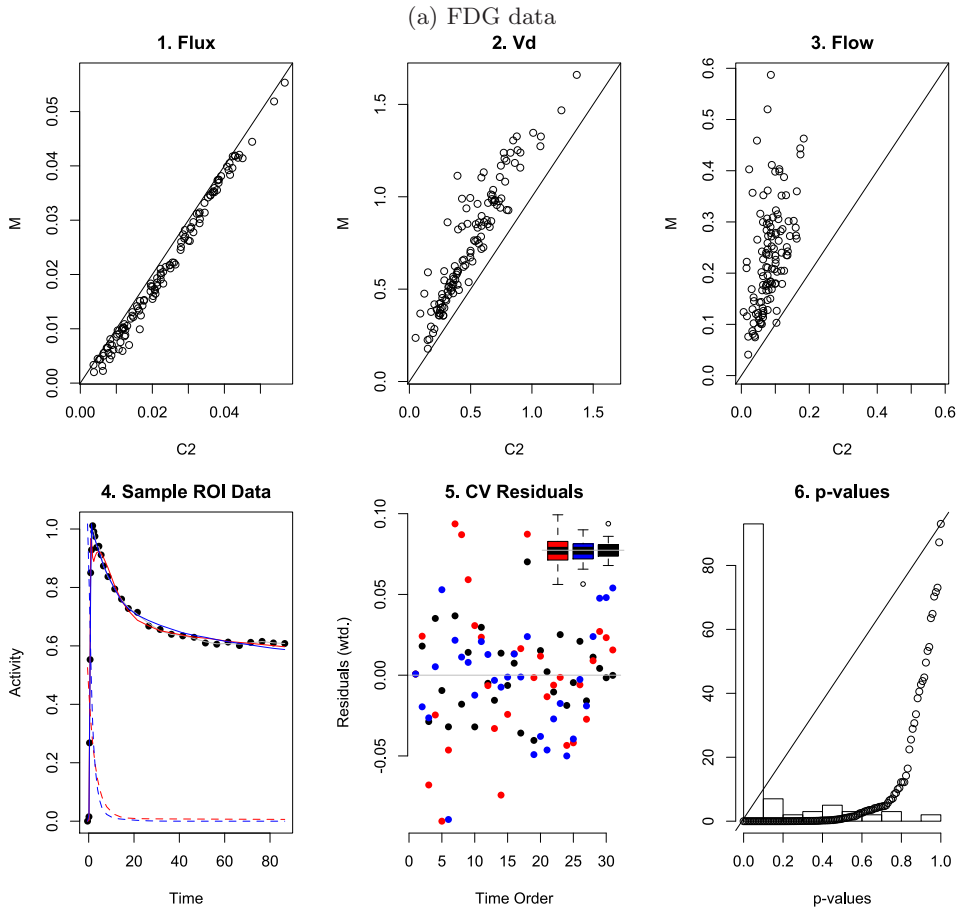}

\end{tabular}
\caption{Comparison between selected kinetic parameters recovered from
analysis of ROI data using compartmental (FDG and H2O) and mixture
models $(1,2,3)$.
4.~Sample ROI data (dots) together with fitted compartment, mixture
and nonparametric analysis (red, blue and grey lines); fitted residues
are shown with broken lines.
5.~Color-coded leave-out-one cross-validated residuals.
6.~Histogram and cumulative distribution (dots) of Wilcoxon $p$-values
for paired comparison of the absolute CV-residuals of the compartmental
and mixture models,
line of identity included.}\label{fig:fig4}
\end{figure}
\setcounter{figure}{3}
\begin{figure}
\centering
\begin{tabular}{@{}c@{}}
\footnotesize{(b) H2O data}\\

\includegraphics{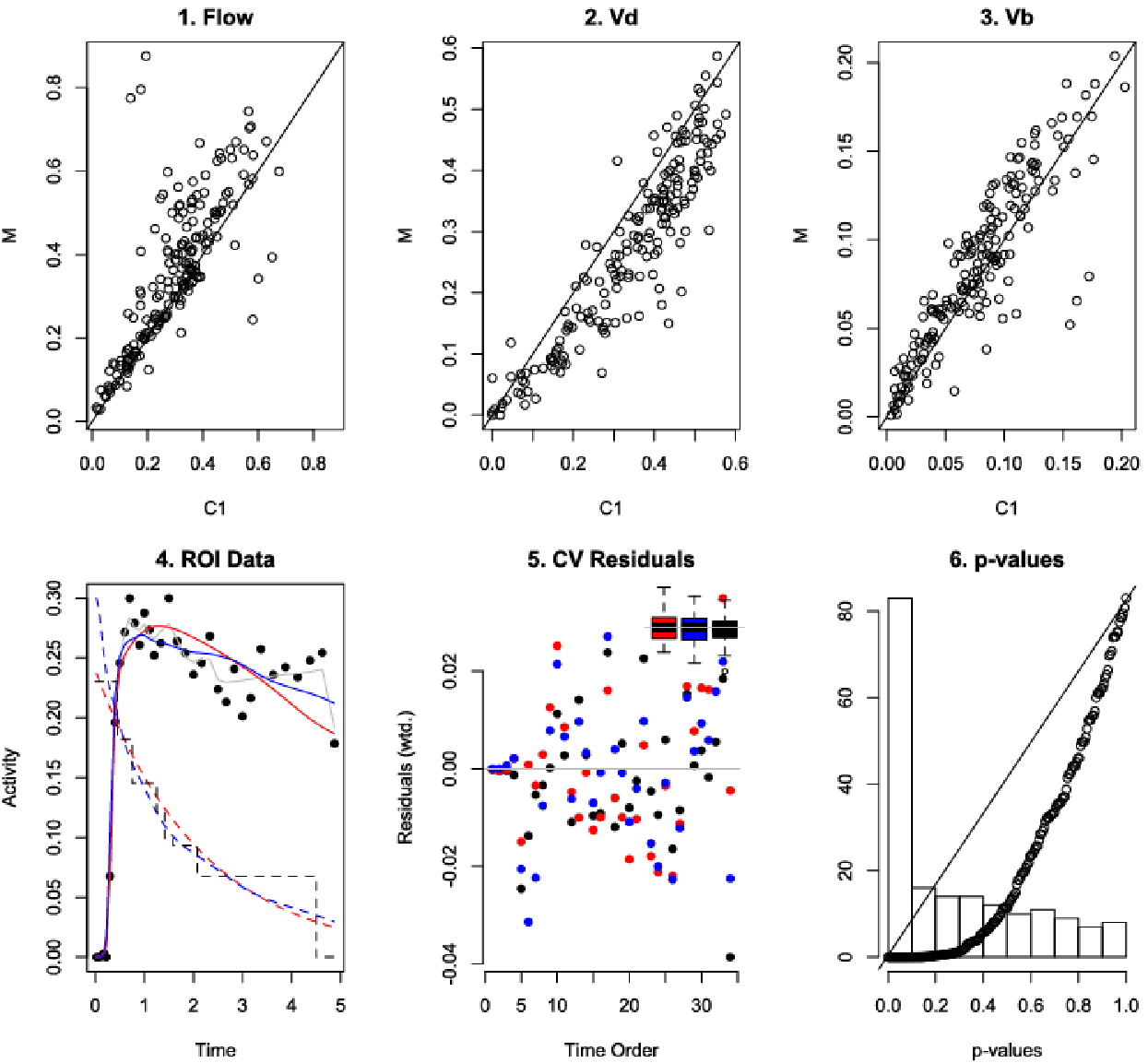}

\end{tabular}
\caption{(Continued).}
\end{figure}

%
%
%
%
%
%
%
%
%
%
%
%
%
%

Results are presented in Figure~\ref{fig:fig4}. With FDG, the cross-validated
residual sum of squares for the adaptive mixture models is
lower than that for the 2-compartment model in 97 of the 120 ROIs
examined (80\% of cases).
This is highly significant.
In addition, the distribution of Wilcoxon $p$-values clearly favor the
mixture model; see Figure~\ref{fig:fig4}(a)6.
Results for H2O are not as strong. Here the adaptive mixture models is
lower than that for the 1-compartmental model in 137 of the 184 ROIs
examined (74\% of cases).
This is again highly significant.
However, the distribution of Wilcoxon $p$-values [Figure~\ref{fig:fig4}(b)6], while
still favoring the mixture model, is more uniform than was found for FDG.
This reduced improvement in the mixture model is very likely a
reflection of the higher noise which is evident in many of the H2O ROIs.
Some representative sample time-course data and the models fit are
also shown in Figures \ref{fig:fig4}(a)4 and \ref{fig:fig4}(b)4.
In the high noise H2O example, the improvement in fit achieved by the
mixture model
is not so clear, however, there is little ambiguity in the FDG case.
Figure~\ref{fig:fig4} also reports comparisons between kinetic parameters recovered
using the mixture and compartmental model
analyses. Note the kinetics from the compartmental and mixture
analysis are quite similar, particularly with flux and volume of distribution
in FDG Figure~\ref{fig:fig4}(a)1--2. Discrepancies in flow values are apparent in
Figure~\ref{fig:fig4}(a)3, with higher flow values being produced by the mixture
analysis model.
Somewhat noisier patterns are found with H2O. Flow and blood volume
values obtained by the
mixture analysis are on average higher than those of the compartmental
model analysis; see Figure~\ref{fig:fig4}(b)1--3.
The 1-compartmental model
produces somewhat higher distribution volume values; see Figure~\ref{fig:fig4}(b)2.
Given the model fit comparisons,
parameters provided by mixture analysis are likely to be more reliable
for FDG and H2O.

%
\subsection{Simulation study}\label{sec5.2}

A prime motivation for mapping of residues at the voxel level is the
simplicity with which the residue for
a region can then be obtained, that is, by simply averaging the
voxel-level residue estimates.
If voxel-level data were measured without error, this approach would be
guaranteed to yield the correct regional residue.
However, one might have concern that
with realistic noise, estimation at the voxel-level might be so poorly
behaved that the regional averaging of voxel-level estimates
would not produce good estimates of the target regional values. Here
the residue recovered from
analysis of the average time course for the region might be more
accurate. We explore this issue in the context
of the brain studies analyzed in Section~\ref{sec4.1}. Recall those analyses
involved consideration of $K=10$ segments, each with a characteristic
time course: $y_k$ for $k=1,2,\ldots,K$.
Each time course was subsequently represented as a linear combination
of a reduced set of ($J$) basis residues
as described in equations (\ref{eq4}) and~(\ref{eq5})---and modeled, similar to
equation (\ref{eq11}), as
\[
y_{kb} = \mu_{kb} + w_{kb}^{-1/2}
\varepsilon_{kb},
\]
where $\mu_{kb} = \sum_{j=1}^J \alpha_{jk} \bar\mu_{jb} (\Delta_k)$,
$w_{kb} = 1/{\mu_{kb}}$ and the
errors $\varepsilon_{kb}$ are found to be approximately
Guassian with mean zero
and constant variance, say, $\phi_k^2$. Given that these data are from
normal subjects,
the configuration of the coefficients $\alpha_{\cdot k} = (\alpha_{1k},\alpha
_{2k},\ldots,\alpha_{Jk}) $ for the $K$ segments is realistic
for normal cerebral PET studies with the FDG and H2O tracers.
We bootstrap from these results to simulate voxel-level data for a set
of $K$ synthetic regions of interest (ROIs).

%
%
%
%
%
%
%
%
%
%

The simulated data for the $i$th voxel in the $k$th region is generated by
%
\begin{equation}\label{eq16}
y_{kib} =\tilde{\mu}_{kib} + \tilde{w}_{kib}^{-1/2}
\tilde\varepsilon_{kib}
\end{equation}
for $i=1,2,\ldots,N_k$ and $b=1,2,\ldots,B$.
Here $\tilde{\mu}_{kib} = \sum_{j=1}^J \tilde\alpha_{ijk} \bar\mu
_{jb}(\tilde\Delta_{ik})$ and $\tilde{w}_{kib}=1/{\tilde\mu_{kib}}$, with
$\tilde\varepsilon_{kib}$, $\tilde\Delta_{ik}$ and $\tilde\alpha_{ijk}
$ independent random variables.
The measurement errors ($\tilde\varepsilon_{kib}$'s) are mean zero
Gaussian with variance $N_k \phi_k^2$ for $i=1,2,\ldots,N_k$.
Amplification of the variance by $N_k$ ensures the mean error over
voxels has variance $\phi_k^2$; the delays ($\tilde\Delta_{ik}$'s) are
log-normal with mean $\Delta_k$ and
standard deviation proportional to the size of the region. The
coefficient of variation is 20\% in the largest region and, finally,
the residue basis scales ($\tilde\alpha_{ijk}$'s) are generated from a
Gamma distribution with
mean $\alpha_{jk}$ and coefficient of variation set in proportion to
the region size. The largest region has a coefficient of variation of
20\%.
This structure is designed to capture the intra-region voxel-to-voxel
variation in terms of
the time-of-arrival of the tracer,
the voxel-to-voxel variation/heterogeneity in residues (including flow)
and, of course, the quasi-Poisson measurement
errors associated with PET instrumentation [Carson et al. (\citeyear{carson})
and \citet{Hues84}].
The overall scale of $\alpha$-coefficients is varied to achieve a range
of six activity levels, equispaced on a logarithmic scale,
the highest of which is
20 times larger than the lowest activity.
The 10 regions had sizes ($N_k$) as follows: 20, 22, 39, 73, 85, 92,
93, 287, 345 and 1519 for FDG;
20, 31, 42, 128, 173, 195, 213, 394, 399 and 925 for H2O.

\begin{figure}
\centering
\begin{tabular}{@{}c@{}}
\footnotesize{(a) FDG simulation}\\

\includegraphics{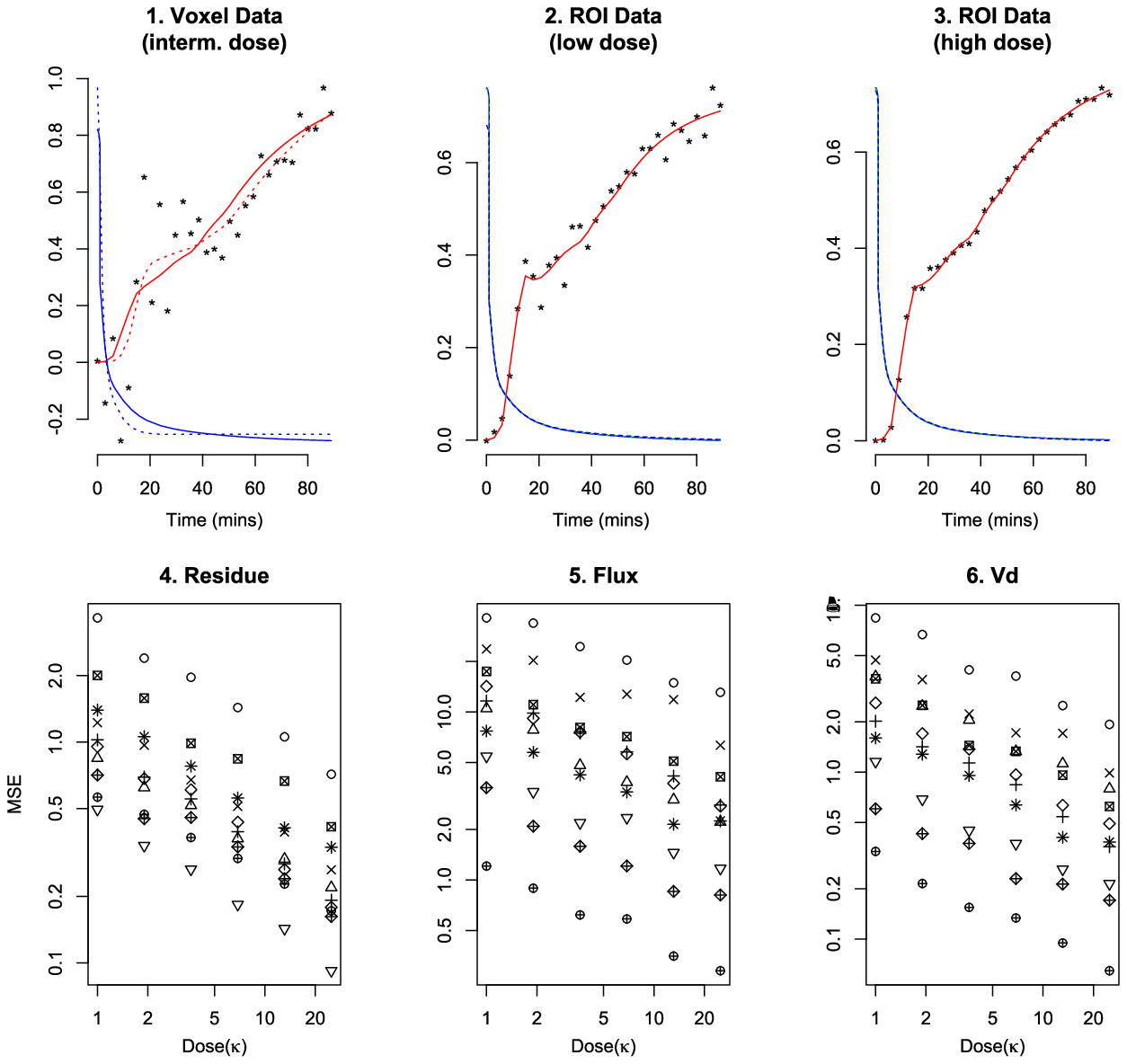}

\end{tabular}
\caption{Sample time activity curve data (dots), together with the true
(solid red) and fitted model (broken red) and the corresponding
residues (blue)---1 (Voxel data), 2 and 3 (ROI data).
Mean square error values (averaged over 50 replicates) for the residue
4 and two other parameters, 5 and 6 (flux and distribution
volume for
FDG and
flow and distribution volume for H2O)
plotted against dose ($\kappa$) for different regions. Regions are
represented by separate symbols.}
\label{fig:fig5}
\end{figure}

\setcounter{figure}{4}
\begin{figure}
\centering
\begin{tabular}{@{}c@{}}
\footnotesize{(b) H2O simulation}\\

\includegraphics{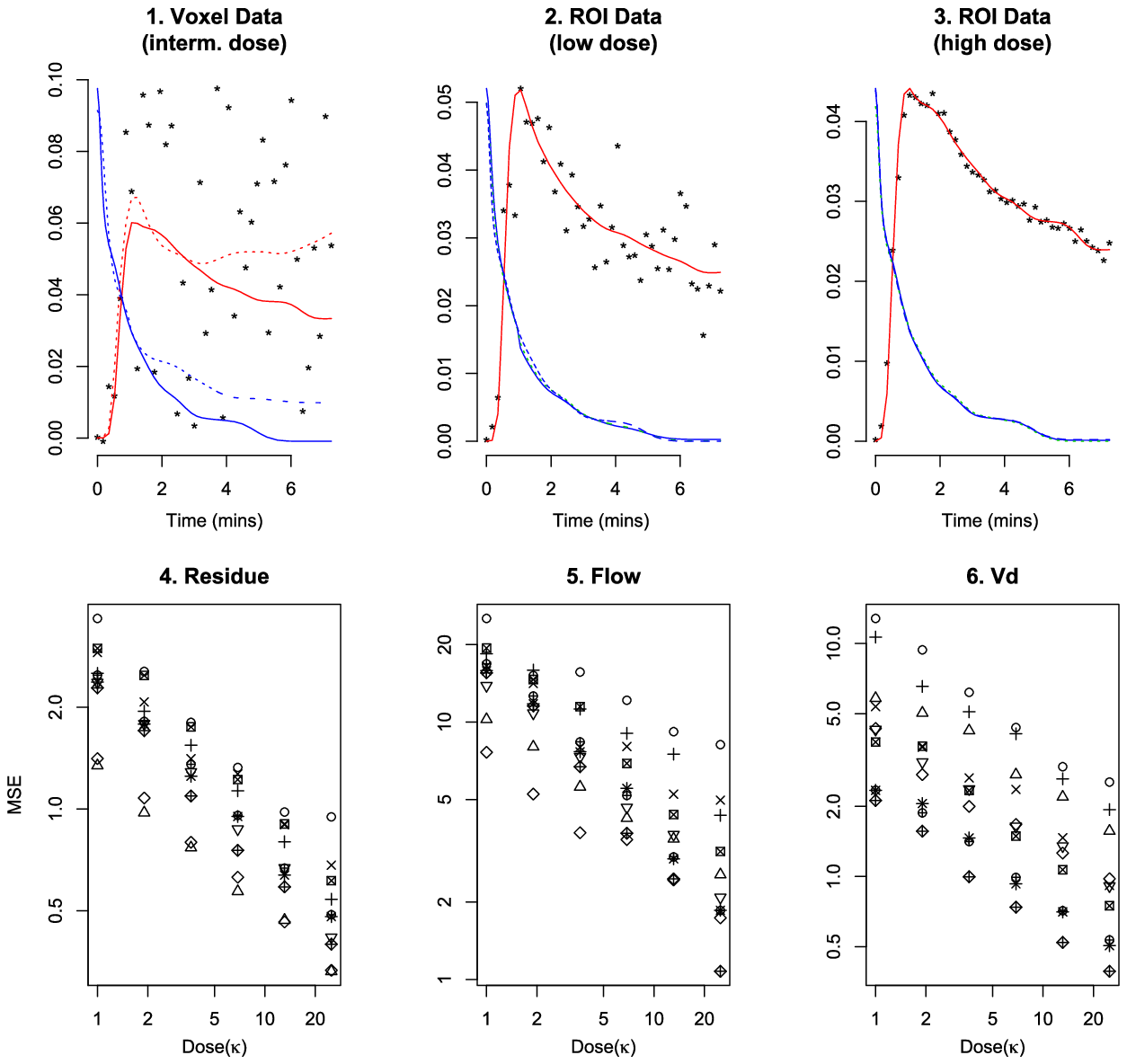}

\end{tabular}
\caption{(Continued).}
\end{figure}

Sample voxel-level and mean ROI time-course data are shown in Figures \ref{fig:fig5}(a)1--3 (FDG) and 5(b)1--3 (H2O).
The intermediate activity/noise data are qualitatively similar to the
data in Section~\ref{sec4.1}.
Sample residue estimates at the voxel and ROI levels are also shown in
these same figures.
Regional residues were computed in 2-ways, one based on analysis of the
mean time-course data for the region and the other
involving averaging of voxel-level residues. Voxel-level analysis could
be done with and without imposition of
positivity constraints on $\alpha$-coefficients.
In high-noise environments,
constrained estimators will be biased. The averaging of such estimators
reinforces the voxel-level bias
leading to a bias in the regional summary. In light of this,
for quantitative regional analysis, the averaging of unconstrained
voxel-level residues is recommended [\citet{huang13}].

Simulations involving data generation for 6 noise levels and 10 regions
were repeated 50 times.
Estimated residues were compared to true values directly using the
integrated squared deviation and also in terms of squared deviation
of key parameters, flux and distribution volume for FDG and flow and
volume of distribution for H2O.
The squared deviations were averaged over replicates to obtain
estimates of
mean square errors (MSE). The MSE for the averaged voxel-level estimators
are plotted against dose in Figure~\ref{fig:fig5}(a)4--6 (FDG) and Figure~\ref{fig:fig5}(b)4--6 (H2O). A very similar pattern is found for the
MSEs estimators produced by the analysis of the regionally averaged
time-course data.
A log-linear pattern is evident in Figure~\ref{fig:fig5}. MSEs vary by region,
with larger regions having smaller MSEs.
An analysis of the MSE data leads to
%
\begin{equation}\label{eq17}
\log(\operatorname{MSE}_{lk}) =\sum_k
\beta_k \cdot I_{lk} - \gamma_a \cdot\log(
\kappa _{lk}) + \gamma_M \cdot M_{lk}+
e_{lk},
\end{equation}
where $\kappa_{lk}$ is dose level, $I_{lk}$ is the region indicator for
region $k$ and $M_{lk}$ is an indicator
of whether the estimator was evaluated by analysis of averaged
time-course data or not.
The modeling error is $e_{lk}$.

\begin{table}
\caption{MSE Characteristics for estimation of regional residues---parameters
defined in the model in equation (\protect\ref{eq17}). $R^2$ values
(adjusted) are from the model
fit}\label{tab1}
\begin{tabular*}{\textwidth}{@{\extracolsep{\fill}}lccc@{}}
\hline
\textbf{Variable} & $\bolds{\gamma_a}$ & $\bolds{\gamma_M}$ & $\bolds{R^2}$ \\
\hline
\multicolumn{4}{c}{FDG}\\
Residue & 0.96 ($\pm0.02$) &$-0.03$ ($\pm0.02$) & 0.99
\\
Flux & 0.88 ($\pm0.02$) & \phantom{$-$}0.05 ($\pm0.02$) & 0.99
\\
$V_D$ & 0.98 ($\pm0.02$) & $-0.01$ ($\pm0.02$) & 0.98
\\ [3pt]
\multicolumn{4}{c}{H2O} \\
Residue & 0.76 ($\pm0.04$) &\phantom{$-$}0.01 ($\pm0.04$) & 0.84
\\
Flow& 0.66 ($\pm0.07$) & \phantom{$-$}0.01 ($\pm0.08$) & 0.98
\\
$V_D$ & 0.95 ($\pm0.02$) & \phantom{$-$}0.01 ($\pm0.02$) & 0.99
\\
\hline
\end{tabular*}
\end{table}

Table~\ref{tab1} reports estimated values for the $\gamma_a$-coefficients.
For FDG the $\hat\gamma_a \approx1.0$, consistent with what one might
expect in theory (see below).
A~slower rate\vadjust{\goodbreak} ($\hat\gamma_a \approx0.8$) is indicated for H2O, a
reflection of the larger number of time frames (42 for H2O versus 31
for FDG) and the greater noise
arising from the rapid decay of the $^{15}$O isotope.
Larger regions generally have lower MSEs, but the heterogeneity of the
region plays a role.
As regards comparison between regional kinetic quantification by
analysis of the averaged time course for the ROI or the averaging of
voxel-level residues, there is little difference.
With H2O there is some small (1\%) degradation in MSE obtained by
analysis of the average time-course data for the region; a similar
result holds for Flux in FDG, although this is not the case for
integrated squared error of the residue or the volume of distribution.
This result supports mapping voxel-level residues because it allows
subsequent analysis of regions of interest to be simply achieved by
direct application of ROIs to
voxel-level residue information.

\subsubsection*{Theoretical interpretation of results}
At high doses MSE can be expected to be dominated by variance.
Since equation (\ref{eq16}) is a linear model, the covariance of the estimated
$\alpha$-coefficients at the $i$th voxel,
which are obtained by a weighted least squares procedure,
will be approximated by
$ \Sigma_{\alpha} \approx\phi^2 [X'WX]^{-1}$,
where
\[
X'WX_{jj'} = \sum_b
w_b \bar\mu_{jb} (\hat\Delta) \bar\mu_{j'b} (\hat
\Delta)
\]
and $w_b = \sum_j \hat\alpha_j \bar\mu_{jb} ( \hat\Delta)$.
Here, to simplify the notation, we have dropped the subscripts $i$ and $k$.
This variance approximation will become more reliable at high doses
when the constraints on the $\alpha$-coefficients are, typically, not active.
Recall from equation (\ref{eq7}),
$\bar\mu_{jb}$ involves the convolution of the normalized residue
$\bar R_j$ and the arterial input function, $C_p$.
So if $C_p$ is expressed as $C_p = \kappa\bar C_p$ where $ \bar C_p$
has amplitude of unity, $ \Sigma_{\alpha} $ can be expressed as
%
\begin{equation}\label{eq18}
\Sigma_{\alpha} = \kappa^{-1} \phi^2 \bigl[\bar
X' \bar W \bar X \bigr]^{-1},
\end{equation}
with $ \bar X' \bar W \bar X$ the same as $X' W X$ but with $\bar\mu
_{jb} $ now involving convolution of the normalized residue and
normalized arterial input $\bar C_p$.
Thus, at high doses
the variance and, consequently, the MSE, will be inversely
proportional to dose ($\kappa$). This is consistent with $\hat\gamma_a
\approx1$, as was found for
FDG in the simulation. The slower convergence ($\hat\gamma_a \approx0.8$)
for H2O may be a reflection of a greater dependence on constraints, due
to the higher noise. When constraints are active, the standard weighted
least squares
covariance formula will not be reliable.

%
\section{Discussion}\label{sec6}

We have presented an approach to the estimation of voxel-level tracer
residues from PET time-course data. The technique uses a data-adaptive
mixture model that allows for voxel-level variation in the time of
arrival of the tracer in the arterial supply.
The mixture representation of local residues
is plausible and has been used previously with basis residues that are
a compartmental model or have simple exponential forms [\citet{cunn-jones}
and \citet{osull93}].
The present work shows that it is possible to also use nonparametric
forms for the basis residues.
This allows the possibility to better investigate potential deviations from
compartmental-like descriptions of tissue residues.
Computationally, the linearity of mixture models is attractive, as it
facilitates the implementation based on efficient use of standard
quadratic programming tools.
The work has a reliance on multivariate statistical methods and uses
backward elimination guided by an unbiased risk type model selection
statistic.\vadjust{\goodbreak}

Residue functions are life tables for the transit time of radiotracer atoms.
Just as infant and elderly mortality patterns might be given separate
attention in a human life table, decomposition of the residue can
provide insight
into the tracer kinetics.
Our approach emphasizes decomposition of residue to focus on flow and
volume characteristics of vascular and in distribution transport
as well as the (apparent) rate of extraction of the tracer by
tissue, that is, flux.
Thus, we have a 5-number summary for the residue.
The life-table perspective on the tissue residue emphasized in this
paper may encourage broader interest in adapting methods from
mainstream survival
analysis for application to the growing needs for quantitation in PET
studies and for related contrast tracking techniques used in
computerized tomography and magnetic resonance [Schmid et al. (\citeyear{Schmidt})].

PET imaging has grown in importance particularly in the context of
cancer, where over 90\% of clinical imaging with PET is carried out.
Having more sophisticated kinetic analysis tools, such as residue
analysis, can enhance the type of information recovered from these studies.
This may potentially lead to better procedures for selecting and
monitoring cancer treatments in order to optimize the patient outcomes.
A number of current clinical imaging trails with PET in cancer already
have reliance on detailed kinetic analysis for extraction of diagnostic
information.
Given the nature of the problems involved, there is an opportunity for
statistics to play a greater role in these developments.

\section*{Acknowledgments} We are grateful to the referees, Associate
Editor and the Editor for a number of comments which led to significant
improvements to the manuscript.




\printaddresses

\end{document}